# Novel hydrogen clathrate hydrate

Yu Wang[1], Konstantin Glazyrin[2], Valery Roizen[3], Artem Oganov[3,4,5], Ivan Chernyshov[6], Xiao Zhang [1], Eran Greenberg[7*], Vitali B. Prakapenka[7], Xue Yang[1], Shu-qing Jiang[1], and Alexander F. Goncharov [1,8]

[1] *Key Laboratory of Materials Physics, Institute of Solid State Physics, Chinese Academy of Sciences, Hefei 230031, Anhui, People's Republic of China*
[2] *Photon Science, Deutsches Elektronen-Synchrotron, Notkestrasse 85, 22607, Hamburg, Germany*
[3] *Moscow Institute of Physics and Technology (State University), Dolgoprudnyi, Moscow region, 141701 Russia*
[4] *Skolkovo Institute of Science and Technology, Skolkovo, 14302 Russia*
[5] *International Center for Materials Discovery, Northwestern Polytechnical University, Xi'an 710072, China*
[6] *TheoMAT Group, ChemBio cluster, ITMO University, Lomonosova 9, St. Petersburg, 191002 (Russia)*
[7] *Center for Advanced Radiations Sources, University of Chicago, Chicago, Illinois 60637, USA*
[8] *Earth and Planets Laboratory, Carnegie Institution of Washington, 5251 Broad Branch Road NW, Washington, DC 20015, USA*
** Current address: Applied Physics Department, Soreq Nuclear Research Center (NRC), Yavne 81800, Israel*

**We report a new hydrogen clathrate hydrate synthesized at 1.2 GPa and 298 K documented by single-crystal X-ray diffraction, Raman spectroscopy, and first-principles calculations. The oxygen sublattice of the new clathrate hydrate matches that of ice II, while hydrogen molecules are in the ring cavities, which results in the trigonal $R3c$ or $R\overline{3}c$ space group (proton ordered or disordered, respectively) and the composition of $(H_2O)_6H_2$. Raman spectroscopy and theoretical calculations reveal a hydrogen disordered nature of the new phase $C_1$', distinct from the well-known ordered $C_1$ clathrate, to which this new structure transforms upon compression and/or cooling. This new clathrate phase can be viewed as a realization of a disordered ice II, unobserved before, in contrast to all other ordered ice structures.**

Water molecules in solid states tend to form hydrogen-bonded 3D frameworks. There are many possibilities for water molecules to associate giving rise to a great variety of configurations and structures observed in nature and at artificial extreme conditions. Pure water is famous for crystallizing in ices of different structures even at moderate pressures resulting in a very complex phase diagram below 2 GPa (*e.g*. Refs. [1, 2] and references therein). This structural diversity is also extended for different water-bearing compounds. Indeed, a variety of clathrate hydrates is known, with water molecules arranging themselves to form polyhedral frameworks with cavities containing foreign guest molecules. The latter interact only weakly with water cages via van der Waals forces. Gas hydrates can be found in many natural environments such as, for example, deep sea sediments and they can play an important role in the planetary formation and evolution [3, 4].

Hydrogen clathrate hydrates have been recently thoroughly investigated at elevated pressures. These materials are of interest for hydrogen storage and are expected to occur in natural environments, icy satellites and comets being the most prominent examples. There are four major crystal structures, which these materials commonly form depending on pressure-temperature (*P*-*T*) conditions, and they differ in structure and composition (Fig. 1). The volumes of cavities in these materials decrease with compression, and the high-pressure structures are commonly described as filled ices – inclusion compounds with structures closely related to known ice structures. The lowest pressure clathrate has a classical sII structure, which forms cages of two different dimensions occupied by a different number of hydrogen molecules [5]. Initially sII structure formation was inferred from the differential thermal analysis data [6], which later were confirmed via *in situ* optical spectroscopy and neutron diffraction investigations [5, 7]. Remarkably, above ~0.35 GPa another hydrogen hydrate phase $C_0$ appears [8-11] consisting of interpenetrating chiral chains of hydrogen-bonded water molecules surrounding orientationally disordered hydrogen molecules in the channels aligned along the ***c***-axis. This structure bears a topological similarity with the mineral quartz and is best described as a filled ice. Above 1.5 GPa, two other hydrogen hydrate phases have been found: $C_1$ and $C_2$ [12]. The structures of these materials resemble those of ices II and VII, respectively [9, 12, 13], so they can be considered as filled ices. It is remarkable that ice II remains the only pure ice modification, which does not have a disordered counterpart [14, 15]. On the other hand, an orientationally disordered $C_2$ filled ice structure was found to be stable above 3 GPa up to at least 80 GPa, although a distortion was reported above 20 GPa [16]. Due to experimental difficulties in assessing structure and chemical composition, especially for hydrogen atoms, an application of first-principles theory is of great benefit in these studies. Recent first-principles calculations confirmed some of the experimental structures and proposed several new, including a high-pressure one ($C_3$) with a very large hydrogen composition (2:1 $H_2$:$H_2O$) above 38 GPa [17].

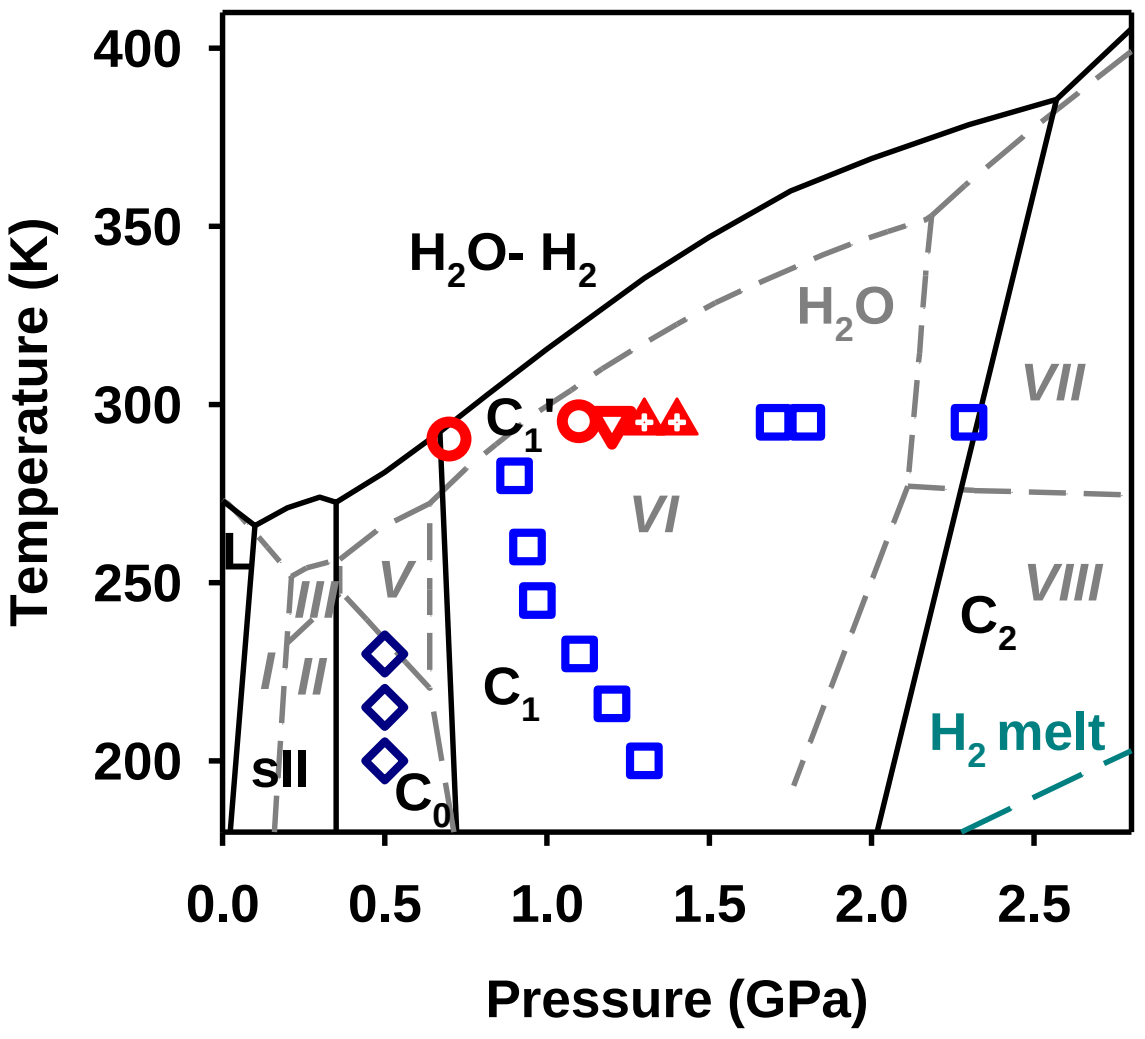

**Figure 1**. Tentative *P-T* phase diagram for $H_2$-$H_2O$ compounds superimposed on those of pure $H_2O$ and $H_2$. Black solid lines are the $H_2$-$H_2O$ phase equilibrium lines proposed in Ref. [9], while gray dashed lines correspond to phase lines for pure water (*e.g*. Ref. [18]) and dark cyan dashed line-for pure $H_2$ [19]. Dark blue open diamonds represent experimentally determined *P-T* conditions in our work for the $C_0$ phase. A new $C_1'$ phase (open red circles and red triangles) appears upon pressure release of the $C_1$ phase at 295 K (open red triangle down) and on pressure increase from fluid (crossed filled red triangles up). The new $C_1'$ clathrate transforms to the known $C_1$ phase (open blue squares) on pressure increase at 295 K and upon cooling down with a simultaneous pressure increase.

In this paper, we report a new hydrogen hydrate (named $C_1'$ here) with a structure, which is very similar to that of $C_1$, but with a disordered hydrogen subsystem within the molecular water framework. This material can be synthesized at about 1.5 GPa at room temperature. We found two possible paths of synthesis: on pressure release of a common $C_1$ clathrate or upon a repeated compression cycle of fluid $H_2O$-$H_2$. This finding emphasizes a tremendous complexity of the structure-composition-pressure phase relationships of hydrogen hydrates and suggests an expansion of the known phase diagram with the implication to hydrogen storage and possible new hydrates in natural environments such as satellites.

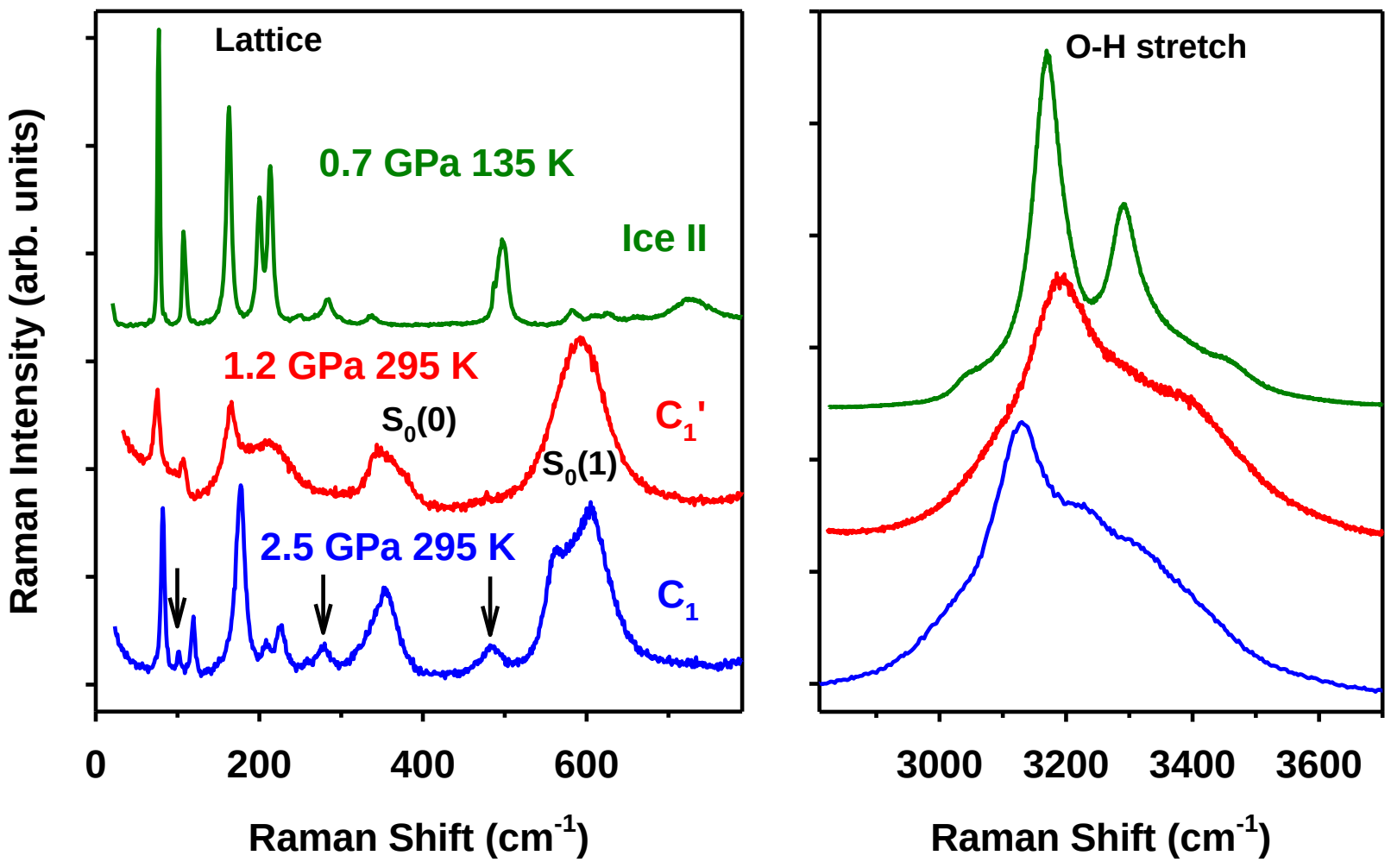

**Figure 2**. Raman spectra of hydrogen hydrates $C_1$ and $C_1'$ and ice II. The *roton* modes of molecular $H_2$ are labeled; the vibron mode at about 4200 $cm^{-1}$ (not shown here) is present for $C_1$ and $C_1'$ and is absent for ice II (Figs. S1, S2). The arrows mark the unique *lattice* modes of $C_1$ phase, which are not observed in $C_1'$ phase, while most of them have counterparts in ice II. Please also note broadening of the lattice modes of $C_1'$ compared to $C_1$. The spectra are shifted vertically for clarity. The vertical scale is different for the left (lattice vibrations) and the right (O-H stretching modes) panels.

The formation of clathrate hydrates is clearly recognized from Raman spectroscopy data, which reveal the incorporation of $H_2$ molecules via the observation of the roton and vibron mode(s) (Figs. S1, S2, Supplemental Material), while characteristic lattice vibration modes indicate the clathrate compound formation and can be used for its identification. At room temperature, we observed the formation of $C_1$ from water-hydrogen fluid mixture upon compression at slightly higher pressures (up to 1.8(1) GPa) compared to that reported previously (*e.g.*, Ref. [12]) most likely due to a higher compression rate. The lattice modes of $C_1$ clathrate are in agreement with the previously reported [10], however, we have extended these measurements below 100 $cm^{-1}$ and observed additional peaks down to 70 $cm^{-1}$ (Fig. 2). A new form of the water hydrogen compound ($C_1'$) occurs upon decompression of the $C_1$ clathrate down to 1.2(1) GPa at room temperature, which can be clearly recognized in the low-frequency range of the lattice vibrations (Fig. S1), where the peaks of $C_1$ phase broaden and some of them disappear (Fig. 2). Despite these distinctions, the lattice modes of $C_1$, $C_1'$, and ice II are all alike suggesting a similar structure of $H_2O$ rings. The transition is reversible by exerting pressure on $C_1'$ phase (Fig. 1). The new $C_1'$ phase also crystallizes upon a repeated compression of a mixed $H_2$+$H_2O$ fluid phase, which occurs on unloading of $C_1'$ phase (Fig. S1). Upon cooling to below 280 K, the $C_1'$ phase transforms into $C_1$ as confirmed by a modification in Raman signal (Fig. S2). Although these observations do not prove thermodynamic stability of $C_1'$ phase, a narrow stability field cannot be ruled out.

Single crystal X-ray diffraction (SCXRD) experiments (Fig. 3) were performed on carefully grown (see Materials and Methods and Fig. S3 in Supplementary Material) and Raman characterized crystals of the $C_1'$ phase, prepared for structure determination. Such measurements have been realized by performing collection of XRD patterns while the sample is rotated along the ω axis (perpendicular to the compression axis) across the aperture of a diamond anvil cell (DAC). During the collection, a series of XRD patterns were acquired for a set of selected ω angles by continuously scanning within 0.5º each 0.5º in a range of ±35º.

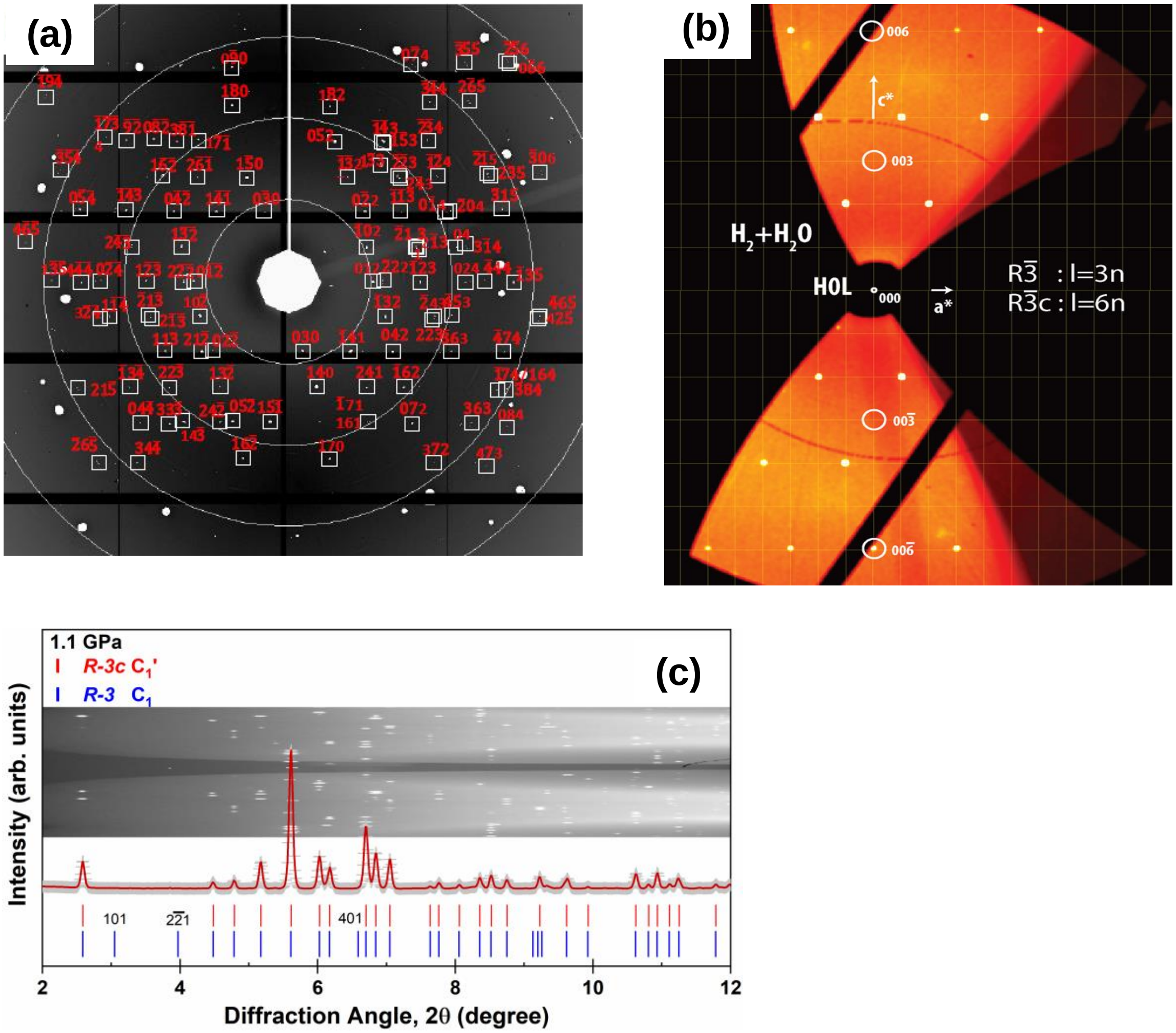

**Figure 3**. (a) SCXRD 2D image collected from a single crystal of $C_1'$ phase at 1.3 GPa upon rotation of ±35º of the DAC demonstrating high quality of a single crystal (Fig. S3); squares mark reflections with the given *hkl* indices; strong unlabeled reflections are from diamond; weak

unlabeled peaks could originate from other smaller single crystal(s), which could not be refined but their contribution to the intensities collected from the large crystal can be definitively singled out by a variety of algorithms (see Supplementary Material); (b) *H0L* slice of reciprocal space visualized by CrysAlisPro software package[20] with attributed diffraction spots. Note the apparent absence of $(003)_{hkl}$ equivalent reflections; (c) Powder XRD data (the X-ray wavelength is 0.291 Å) integrated in 1D pattern (gray plus symbols present the data and red solid line is the Le Bail fit), the short red ticks correspond to $\boldsymbol{R\bar{3}c}$ space group with lattice parameters a = 13.102(4) (Å), b = 6.263(7) (Å). The blue ticks also represent XRD of the $C_1$ $\boldsymbol{R\bar{3}}$ structure (the same space group as ice II) with the same lattice parameters. Inset shows an image of the 2D integrated powder XRD data of $C_1'$ phase.

Structural information on the $C_1'$ phase was obtained from indexing the collected single crystal data set (Fig. 3(a)) and the subsequent structure solution (Table 1). Analyzing 1D powder XRD patterns (Fig. 3(c)) is not sufficient to distinguish this phase from $C_1$ and ice II with the same lattice parameters. Indeed, our powder XRD measurements were not able to detect the $C_1'$ to $C_1$ phase transition upon pressure increase at 295 K (Fig. S4) because the difference is very subtle, making our single-crystal results on $C_1'$ critical for identification of this phase. Ice II, which is stable below 0.7(1) GPa (Fig. 1), has very different unit cell parameters and density than $C_1'$ and $C_1$ phases (Fig. S5), so it can be ruled out.

Indexing of the observed diffraction Bragg spots of $C_1'$ phase suggested presence of trigonal symmetry. By analyzing the reflection rules for the observed diffraction spots, we can clearly rule out the rhombohedral $\boldsymbol{R\bar{3}}$ structure of ice-II as illustrated in Fig. 3(b), which represents a slice of reciprocal space for the $C_1'$ crystal. Here we report that $(003)_{hkl}$ as well as its equivalent reflection $(00\bar{3})_{hkl}$ are evidently absent, while $(006)_{hkl}$ is indeed present. This observation allowed us to exclude $\boldsymbol{R3}, \boldsymbol{R3m}, \boldsymbol{R\bar{3}m}$ in a similar fashion and select $\boldsymbol{R3c}$ and $\boldsymbol{R\bar{3}c}$ space groups as the most probable candidates. This result also shows a clear distinction between the structure of the new $C_1'$ clathrate in comparison to that of $C_1$, which has the same $\boldsymbol{R\bar{3}}$ symmetry as ice-II [12].

The individual Bragg peak intensities were extracted by means of Rigaku CrysAlisPro software [20] and analyzed by means of Olex2 [21] (with SHELX[22] backend) and additionally with Jana2006 [23]. During the structure solution, the oxygen thermal displacement parameters were refined as anisotropic ($U_{anis}$), while isotropic approximation was used for hydrogen atoms ($U_{iso}$). Electronic density corresponding to $H_2$ was located in the center of the framework cavities, but due to the limited resolution and averaging property of the XRD signal, we could not resolve position of individual H atoms and assigned $H_2$ molecule to a single position. The positions of the hydrogen atoms bound to the oxygen rings were found and refined in our analysis.

As shown in the Table 1, structure solutions within the space groups $\boldsymbol{R3c}$ and $\boldsymbol{R\bar{3}c}$ describe the experimental results almost equally well. In both of these structures, linear arrays of $H_2$ molecules are confined in channels made of the $H_2O$ rings. The difference between these two structural solutions is very subtle because of the weak scattering contribution of the hydrogen atoms. While the positions of the oxygen atoms are well constrained, there is an important difference between the solutions concerning the positions and populations of the hydrogen atom sites. Depending on the solution, the hydrogen atoms of the $H_2O$ rings are ordered for $\boldsymbol{R3c}$ and disordered for $\boldsymbol{R\bar{3}c}$ space groups. A simple addition of a symmetry element will break the spatial confinement of H atoms with respect to the individual O atoms (as attributed to $\boldsymbol{R3c}$) and allow a random disorder of individual H atoms between the equivalent general crystallographic positions of $\boldsymbol{R\bar{3}c}$. These half-populated crystallographic positions are lying close the lines between the O-O atoms forming the rings or between the oxygen positions in the adjacent rings shifted along the $\boldsymbol{c}$ axis (Fig. 5). The detailed crystallographic description of our structural $\boldsymbol{R3c}$ and $\boldsymbol{R\bar{3}c}$ solutions along with the experimental and fitted reflection intensities is presented in the Supplementary Material in the form of the standard structural files (cif).

Distinguishing between the $\boldsymbol{R3c}$ and $\boldsymbol{R\bar{3}c}$ within the given XRD dataset is not straightforward, due to the limitations imposed by high-pressure environment. Considering the data from the statistics point of view (Table 1), we see that the $\boldsymbol{R3c}$ R-factor is not significantly lower than the one of $\boldsymbol{R\bar{3}c,}$ although the number of fit parameters has increased by ~40%. This could indicate that the assignment to $\boldsymbol{R\bar{3}c}$ is indeed the appropriate one [24]. Additional arguments for the disordered nature of $C_1'$ phase can be provided by Raman spectroscopy and first-principles calculations.

**Table 1.** Lattice parameters and refinement indicators of the experimental single crystal XRD measurements at 1.3(1) GPa (see supplementary information for the full set of data including the atomic positions).

| Space group | Lattice parameters a, c (Å) | Chemical formula | $N_{obs}/N_{par}$ | $R_{F2}$ factor | Obs. HKL | Residual electronic density* | |
|---|---|---|---|---|---|---|---|
| | | | | | | neg. | pos. |
| $\boldsymbol{R\bar{3}c}$, #167 | 12.984(4), 6.243(8) | $(H_2O)_{36}(H_2)_6$ | 130/**23** | 2.5% | $1 < H < 17$<br>$-6 < K < 0$<br>$-7 < L < 8$ | -0.1 | +0.1 |
| $\boldsymbol{R3c}$, #161 | 12.984(4), 6.243(8) | $(H_2O)_{36}(H_2)_6$ | 130/**33** | 2.0% | | | |

$N_{obs}$ – number of observed reflections with I > 3•sig(I), where I is the intensity of a Bragg spot, and sig(I) is equal to the square root of intensity.
$N_{par}$ – number of refinement parameters
$R_{F2}$ – crystallographic R-factor obtained on refinement using intensities, or structure factor squared
* – as reported by SHELXT, part of SHELX package.

Raman mode group theory analysis reveals a different activity for the ordered and disordered candidate structures of $C_1'$ (Table S1). This is because a disordered $\boldsymbol{R\bar{3}c}$ structure has only one distinct oxygen position and it is centrosymmetrical, making only the even symmetry (gerade) modes Raman active. Raman active modes of the $C_1$ structure are also subjected to this rule, but, nevertheless, this structure shows more Raman activity compared to the $\boldsymbol{R\bar{3}c}$ $C_1'$ because it is less symmetrical. Our experiment shows less Raman activity of $C_1'$ compared to $C_1$, thus clearly supporting a disordered $\boldsymbol{R\bar{3}c}$ $C_1'$. In addition, a broadening of the $C_1'$ Raman peaks (Fig. 2) indicates its disordered structure.

To give an independent insight into the structure and composition of $C_1'$ phase, we performed the state of the art *ab initio* calculations mainly focusing on the candidate structures $\boldsymbol{R3c}$ and $\boldsymbol{R\bar{3}c}$.

We examined the stability of newly synthesized $C_1$' phase using the first-principles DFT calculations. First, we performed structure search (employing USPEX [25]) at the pressure of 1.4 GPa. The structure search by USPEX algorithm predicted an orientationally ordered structure with a $\boldsymbol{R3c}$ space group (see Table S2). The transition pressure from the ordered approximation of $C_1$' to $C_1$ clathrate structure is predicted to be ~2.1 GPa (see Fig. S6), which is broadly consistent with the experiment.

Our next step was to investigate a structural disorder of hydrogen atoms, which was conducted by analyzing all O(H)···O contacts. These appear to be split in two groups: “cyclic” inside $(H_2O)_6$ rings and “acyclic” between the rings bounding them together (Fig. 4). The randomly generated disordered structures that include the disorder of both kinds and their mixture were relaxed by VASP and their enthalpies were calculated (see Materials and Methods and Figs. S6-S7). The computations show that these disorders become energetically favorable at different temperatures: cyclic at 286 K, acyclic at 68 K and the combination of them at 287 K, which is again broadly in agreement with the experiment suggesting the full proton disorder in the new clathrate structure. The structure of $C_1$' phase that includes both types of disorder suggests the occupancy of hydrogen positions at the rings to be 0.5, which agrees well with the experiments (structure $\boldsymbol{R\bar{3}c}$ of the Table 1 and Table S2).

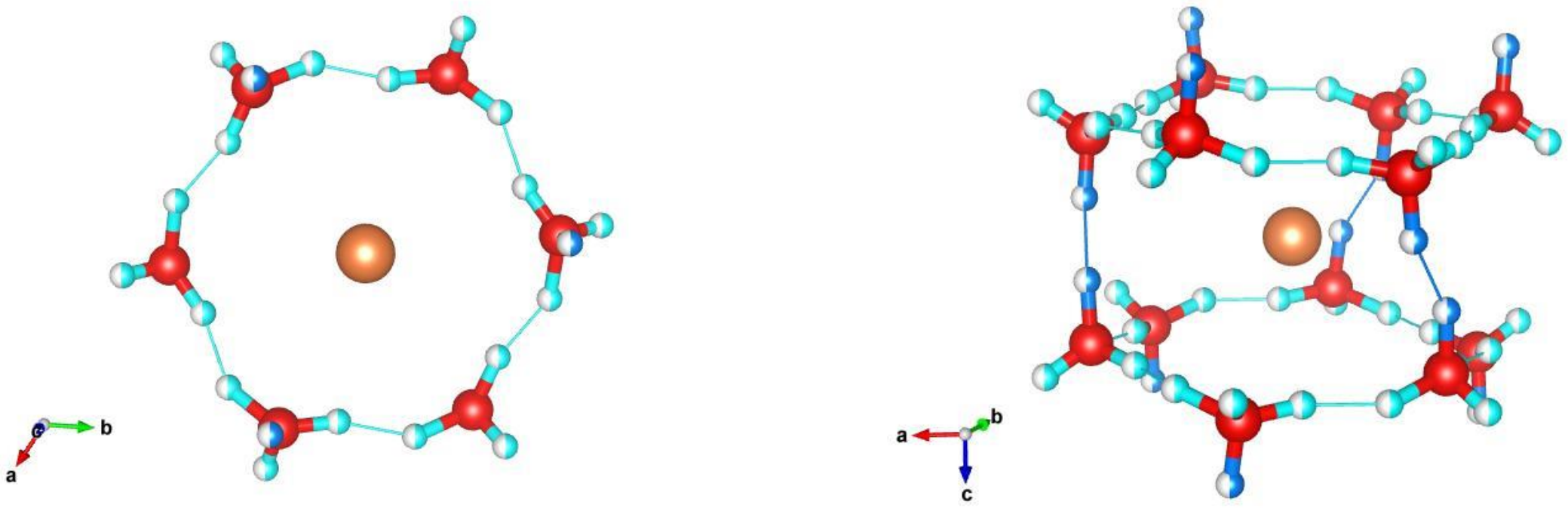

**Figure 4**. The structure of $C_1$' clathrate in $\boldsymbol{R\bar{3}c}$ setting, where the major structural motif ($H_2O$ rings) is shown in different projections to demonstrate two different disordering schemes. Orange spheres correspond to the centers of mass of the hydrogen molecules. Bluish spheres of different shades correspond to partially occupied positions for hydrogen atoms in the water cages. The solid thin light cyan and blue lines connect the duplicated hydrogen positions, generating “cyclic” (in an armchair-like virtual plane formed by six oxygen atoms) and “acyclic” (outside the plane) types of disorder, respectively.

Our Raman data of $C_1$' phase strongly point to a spatially disordered $\boldsymbol{R\bar{3}c}$ structure of the $H_2O$ rings because both the lattice and OH-stretch bands are broader and a number of lattice modes are missing if compared to those of ordered $C_1$ phase (Fig. 2, Table S1). On the other hand, the most prominent lattice mode peaks at 90 and 160 $cm^{-1}$ nearly coincide in frequency for $C_1$ and $C_1$' phases indicating the proximity of their structures. The comparison of the experimental and theoretical (for the ordered structure) Raman spectra shows a good correspondence (Fig. S8) in a number and the spectral positions of the peaks. The lattice modes are slightly lower in frequency than the theoretically computed, and the peak broadening caused by anharmonicity or disorder varies in experiment for different peaks, while was assumed constant in calculations. The discrepancy in the position of the O-H stretching modes for $C_1$' phase is at least partially due to the sensitivity of these modes to the unit cell volume (or pressure) (see Table S3).

The chemical composition of the new $C_1$' clathrate is the same as of $C_1$ clathrate determined by combining SCXRD structural solution and DFT calculations. This results in the $(H_2O)_6H_2$ composition. At 1.3 GPa the density is 1.20(1) g/cm$^3$. This can be compared to the density calculated from the molecular volumes of 23.6 Å$^3$/$H_2O$ for ice II [26] and *hcp* phase I hydrogen [27] with 3.1 Å$^3$ per hydrogen atom (Fig. S5). The compound is some 5.5% denser than the mechanical mixture of the latter ice and hydrogen, which points to an exceptional stability of clathrate hydrates at these P-T conditions, which is further emphasized by their high melting temperature that is higher than in pure ice and hydrogen (Fig. 1). The $C_1$' clathrate can be considered as a disordered $C_1$, with respect to which it is likely stable given the conditions of its formation.

In conclusion, our combined SCXRD, Raman spectroscopy, and first-principles calculations demonstrate the existence a new hydrogen clathrate hydrate can be synthesized at 1.2 GPa and 298 K via pressure manipulation. The composition of this compound is $(H_2O)_6H_2$, the same as of the familiar $C_1$ clathrate, while the structure is similar to $C_1$ clathrate and ice II but proton-disordered. This is remarkable given that a disordered version of ice II is not known. This findings shed new light on the phase diagram of water and clathrates. Future neutron single crystal diffraction investigation would be instrumental in confirmation of these results.

# Supplementary Material

# Novel hydrogen clathrate hydrate

Yu Wang [1,2], Konstantin Glazyrin[3], Valery Roizen[4], Artem Oganov[4,5,6], Ivan Chernyshov[7], Xiao Zhang [1,2], Eran Greenberg[8*], Vitali B. Prakapenka[8], Xue Yang[1], Shu-qing Jiang[1], and Alexander F. Goncharov [1,2,9]

[1] *Key Laboratory of Materials Physics, Institute of Solid State Physics, Chinese Academy of Sciences, Hefei 230031, Anhui, People's Republic of China*
[2] *University of Science and Technology of China, Hefei 230026, Anhui, People's Republic of China*
[3] *Photon Science, Deutsches Elektronen-Synchrotron, Notkestrasse 85, 22607, Hamburg, Germany*
[4] *Moscow Institute of Physics and Technology (State University), Dolgoprudnyi, Moscow region, 141701 Russia*
*Skolkovo Institute of Science and Technology, Skolkovo, 14302 Russia*
[6] *International Center for Materials Discovery, Northwestern Polytechnical University, Xi'an 710072, China*
[7] *TheoMAT Group, ChemBio cluster, ITMO University, Lomonosova 9, St. Petersburg, 191002 (Russia)*
[8] *Center for Advanced Radiations Sources, University of Chicago, Chicago, Illinois 60637, USA*
[9] *Earth and Planets Laboratory, Carnegie Institution of Washington, 5251 Broad Branch Road NW, Washington, DC 20015, USA*
** Current address: Applied Physics Department, Soreq Nuclear Research Center (NRC), Yavne 81800, Israel*

This pdf file contains Materials and Methods, Supplementary Figures S1-S8, Tables S1-S3, and Bibliography with 18 References.

Two cif files corresponding to the two main structural models for $C_1'$ structure at 1.3 GPa of this work are presented separately.

Corresponding author:

agoncharov@carnegiescience.edu

## MATERIALS and METHODS

### Experiments

We performed the experiments in a modified symmetric diamond anvil cell (DAC) with culets of 500 µm in diameter equipped with stainless steel or rhenium gaskets drilled with a 150 µm hole. A small vesicle of doubly distilled and deionized water was loaded in the cavity leaving a space for air, which was misplaced by compressed ultrapure $H_2$ gas at about 0.2 GPa using a high pressure gas loading vessel at room temperature. A powder $C_1$ clathrate sample forms by compressing the mixture of hydrogen and water, while powder $C_1'$ samples originate from the unloaded $C_1$ phase. Single crystals of the $C_1'$ phase (see supplementary information, Fig. S3) were grown by cycling pressure near the crystallization transition, where small spontaneously emerging crystals of $C_1'$ phase recrystallized to form to a larger single crystal extending across the whole high-pressure cavity of the DAC. All measurements were performed on samples with an excess of hydrogen to avoid the formation of pure ice phases. The pressures were determined using the $R_1$ line spectral position of the ruby fluorescence[1].

Four sets of independent experiments were performed probing the formation of different clathrates by Raman spectroscopy. Another three sets of experiments were conducted on both powder X-ray diffractions (XRD) and single crystal x-ray diffraction (SCXRD) to identify the structure of a new phase, preselected by Raman spectroscopy, that show a distinct spectrum attributed here to $C_1'$ clathrate. Our confocal Raman probe operating at room and low temperatures uses either of three laser excitation wavelengths 488, 532, and 660 nm, which are delivered by narrow-line solid-state lasers. The Raman notch filters (three per each excitation wavelength) are of a very narrow bandpass (BragGrate™ Optigrate©) allowing Raman measurements down to 10 $cm^{-1}$ in the Stokes and anti-Stokes. One of these notch filters is used as a beamsplitter to inject the laser beam into the optical path. A custom-made microscope with a Mitutoyo 20X long-working distance objective lens collects the Raman signal in a back scattering geometry. A single stage imaging spectrograph SP2500 (Acton) with a 500 mm focal length was equipped with an array thermoelectrically cooled CCD detector (Princeton eXcelon). Normally, spectra were acquired in a low resolution mode (300 grooves/mm grating) for a quick reconnaissance followed by a longer time (<300 s) acquisition using a high spectral resolution mode (1200 grooves/mm grating) yielding approximately a 4 $cm^{-1}$ spectral resolution. The laser power was limited to values below 10 mW at the sample to avoid sample overheating. The advantage of Raman spectroscopy, appreciated by the community, and experimentally proven on many systems, is that it provides a robust initial diagnostics of a new phase formation allowing performing XRD on well-characterized samples.

Single crystal synchrotron XRD experiments were performed at room temperature by collecting patterns at different angle positions of a DAC along the ω rotation axis (perpendicular to the X-ray beam and the compression axis). The samples were carefully centered on a rotation axis. The opening angle of the used four-pin symmetric and BX90 DACs was ±35º (X-ray aperture). The XRD patterns were acquired with a 0.5º interval with a collection time of 1-4 seconds at each point. XRD experiments were performed several times on two beamlines: GSECARS sector 13,

Advanced Photon Source (APS), Argonne National Laboratory with wavelength of 0.3344 Å, and PETRA III Deutsches Elektronen-Synchrotron (DESY) with wavelength of 0.291 Å (preliminary powder data) yielding quite consistent results.

**Theoretical calculations**

The stability of the newly discovered $C_1$' phase was investigated by *ab initio* simulations. We started from the crystal structure prediction of the ordered version of the $C_1$' phase using the evolutionary algorithm USPEX [2-5]. The fixed composition molecular calculation was done at 1.4 GPa with following settings. There were 80 random structures in the first generation, and the next generations consisted of 60 structures produced by 40% of random structures and 40%, 10%, 10% made by heredity, rotational mutation and lattice mutation respectively.

All structure relaxations and total energy calculations were performed using density functional theory (DFT) as implemented in the Vienna ab initio simulation package (VASP) [6]. Projector-augmented wave (PAW) [7] GW-potentials were used (with He [$2s^2 2p^4$] core in case of oxygen). We used a plane-wave energy cutoff of 700 eV and a Γ-centered k-point mesh with a resolution of $2\pi \times 0.04$ Å$^{-1}$. The many-body dispersion energy method was used for vdW corrections [8, 9]. To determine the transition pressure from $C_1$' to $C_1$ phase we conducted *ab initio* calculations with the same settings in a pressure range from 1 to 3 GPa with steps of 0.2 GPa.

In order to compare experimental and theoretical results we performed Raman intensity calculations based on DFT [10] as implemented in the Fonari and Stauffer script [11] for processing of the results of VASP calculations. For these simulations the non-local correlation functional of Langreth and Lundqvist et al. were used [12-14]. Prior to these simulations all structures were fully relaxed until all force components became less than 0.01 meV / Å in absolute value. For calculations of Raman spectra, the hexagonal representations of the trigonal unit cells were used.

In addition, the calculation of structural disorder was conducted. First, we estimated the difference between the average enthalpy of the series of disordered structures and the enthalpy of the USPEX-predicted structure. Then, by dividing these differences by disorder entropy in accordance with Pauling approximation [15] we were able to calculate disorder temperatures for three types of disorder.

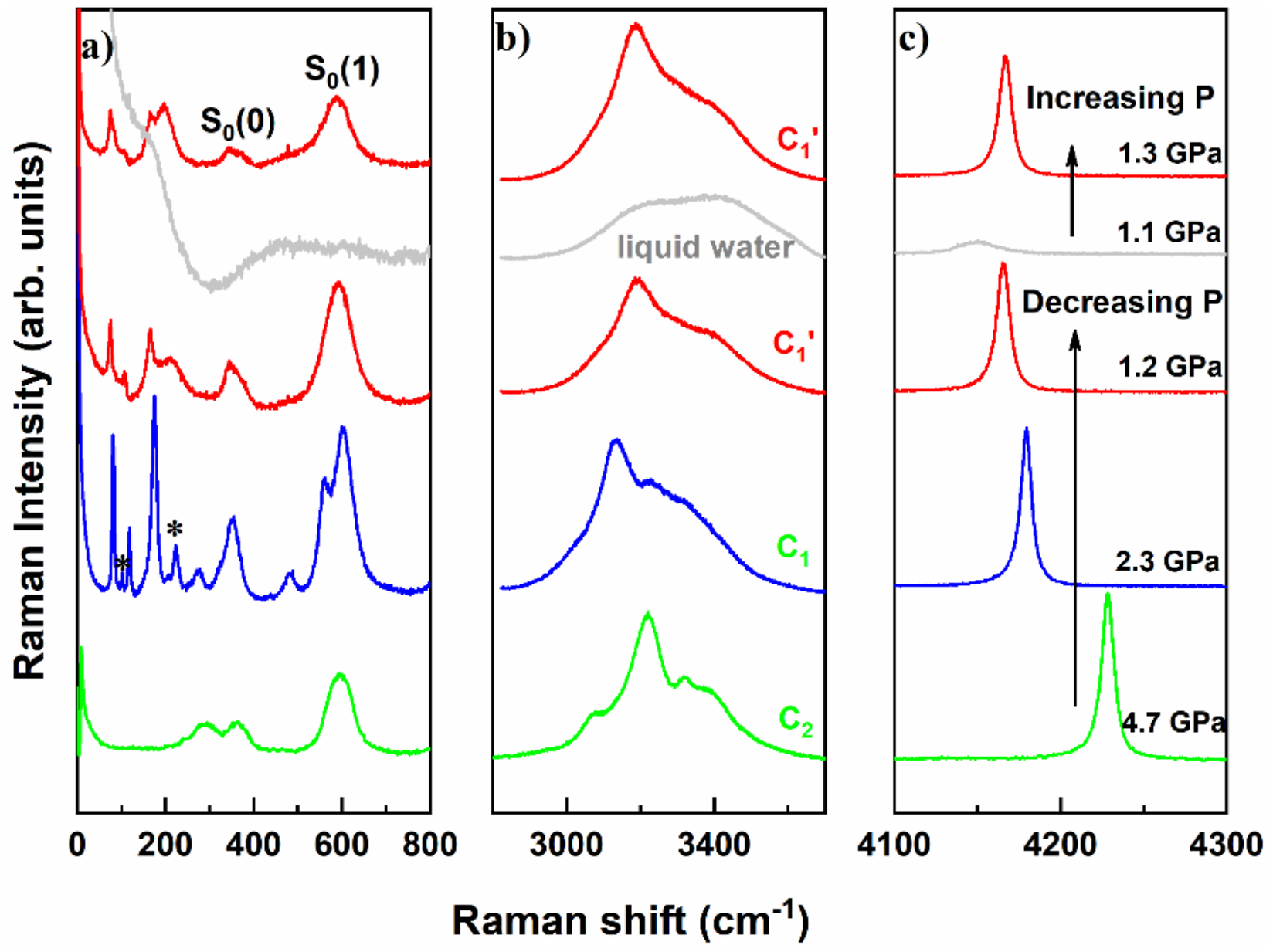

**Figure S1**. Raman spectra of hydrogen hydrates measured upon decompression inducing the formation of $C_1'$, followed by re-compression. Panels (a, b, c) correspond to the regions of the specific *lattice* vibrations, O-H stretching, and $H_2$ *vibron* modes, respectively. Asterisks emphasize the changes of the lattice mode due to the phase transition from phase $C_1$ to $C_1'$. The peaks marked as $S_0(0)$ and $S_0(1)$ are the strongest *roton* modes of hydrogen molecules. The arrows represent the sequence of performing experiments. The intensity scale in each panel is different. The spectra are shifted vertically for clarity. The vertical scale is different for all the panels.

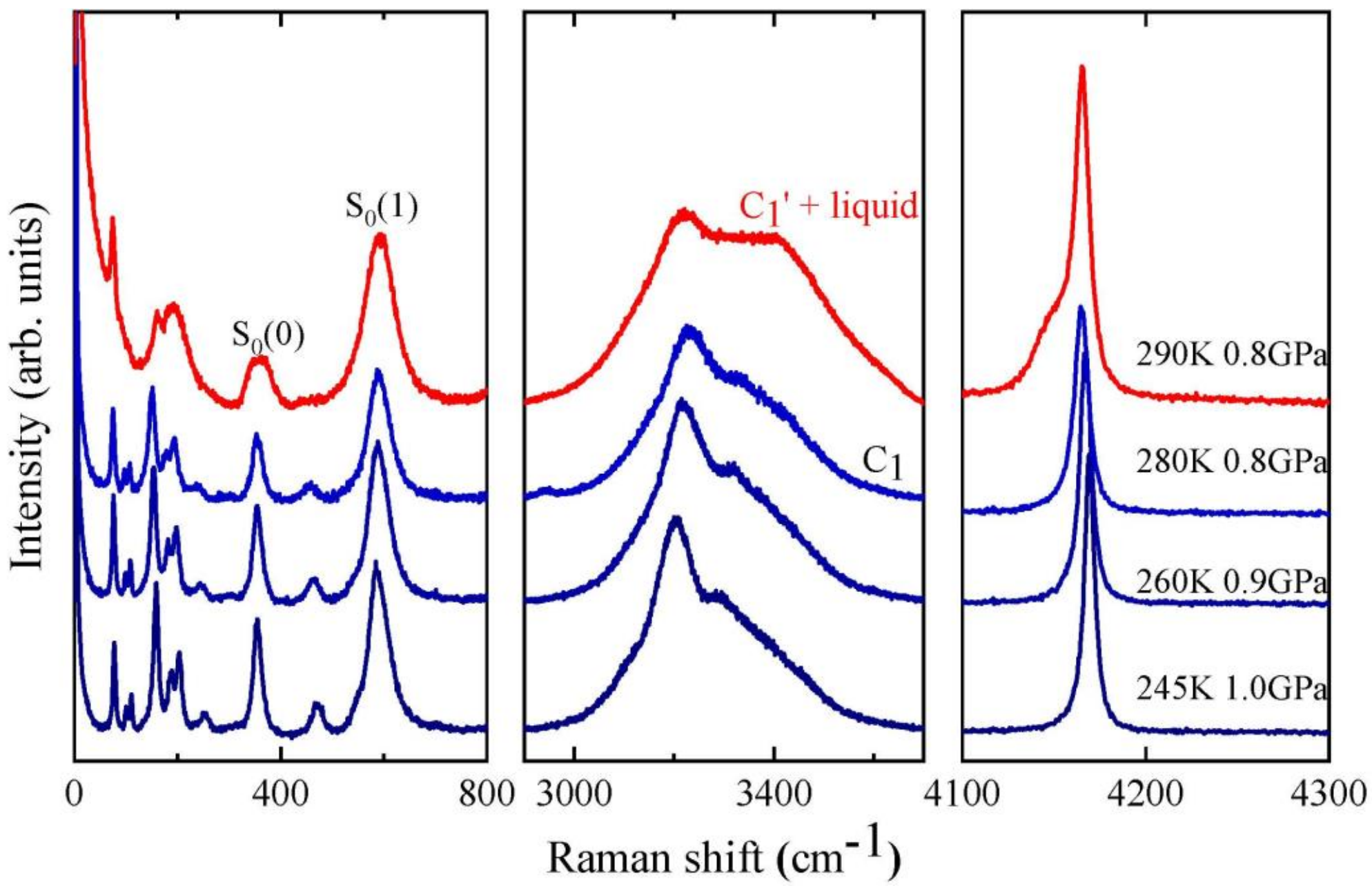

**Figure S2**. Raman spectra of $C_1$' and $C_1$ clathrates upon cooling down.

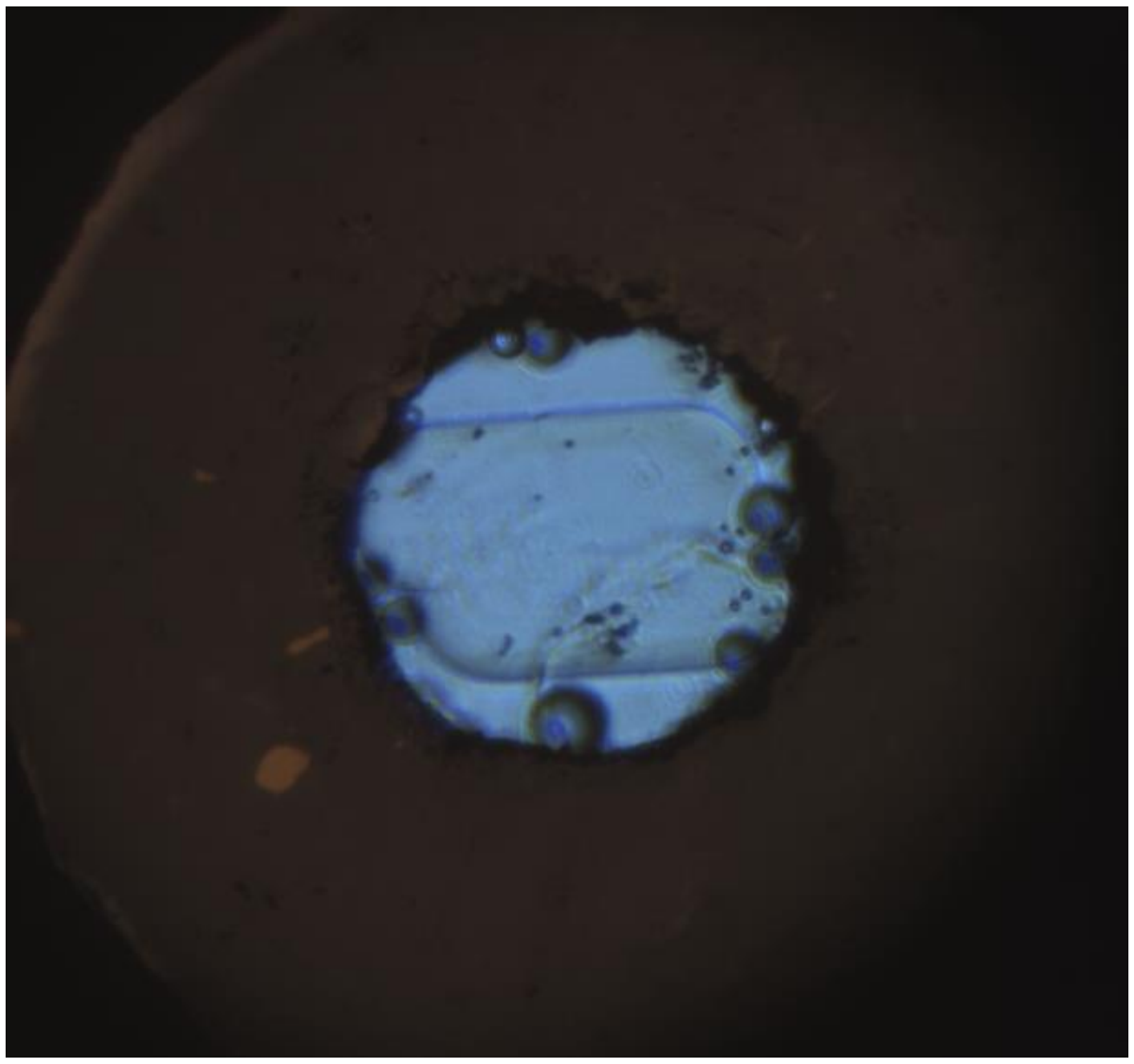

**Figure S3**. A microphotograph of a $C_1$' single crystal at 1.5 GPa, which was used in SCXRD measurements. The crystal in the middle of the cavity is surrounded by $H_2$ medium, where ruby balls are immersed for the pressure measurements. The culet is 300 μm and the cavity is about 100 μm in diameter.

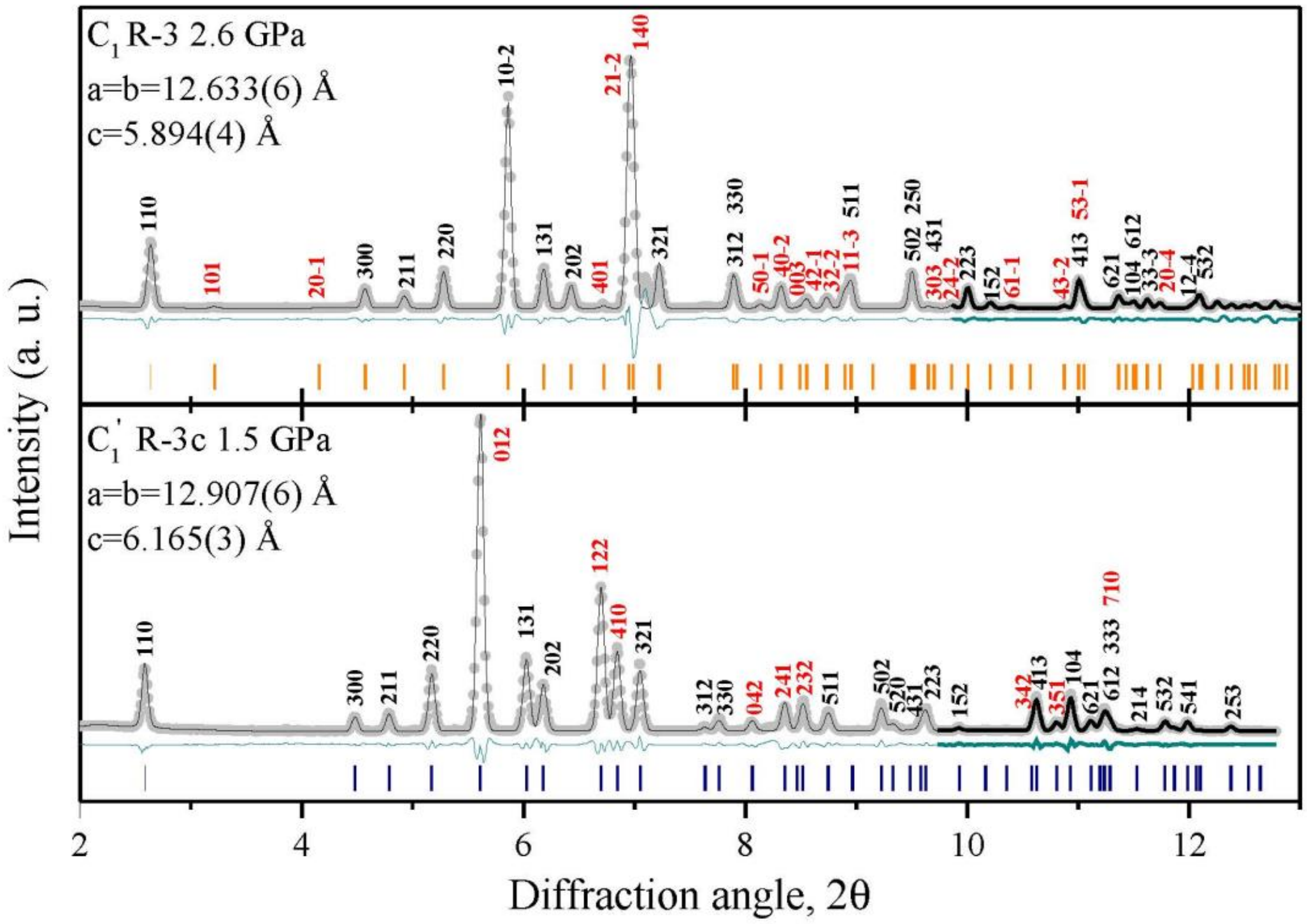

**Figure S4.** Powder XRD data (the X-ray wavelength is 0.291 Å) for $C_1$ and $C_1'$ phases (identified by Raman spectroscopy, Fig. S1) integrated in 1D patterns (gray filled circles present the data and black solid line is the Le Bail fit); the short blue ticks correspond to $\boldsymbol{R\overline{3}c}$ space group of $C_1'$ and the yellow ticks – $\boldsymbol{R\overline{3}}$ structure of $C_1$ phases, respectively, with the refined lattice parameters presented in the legends.

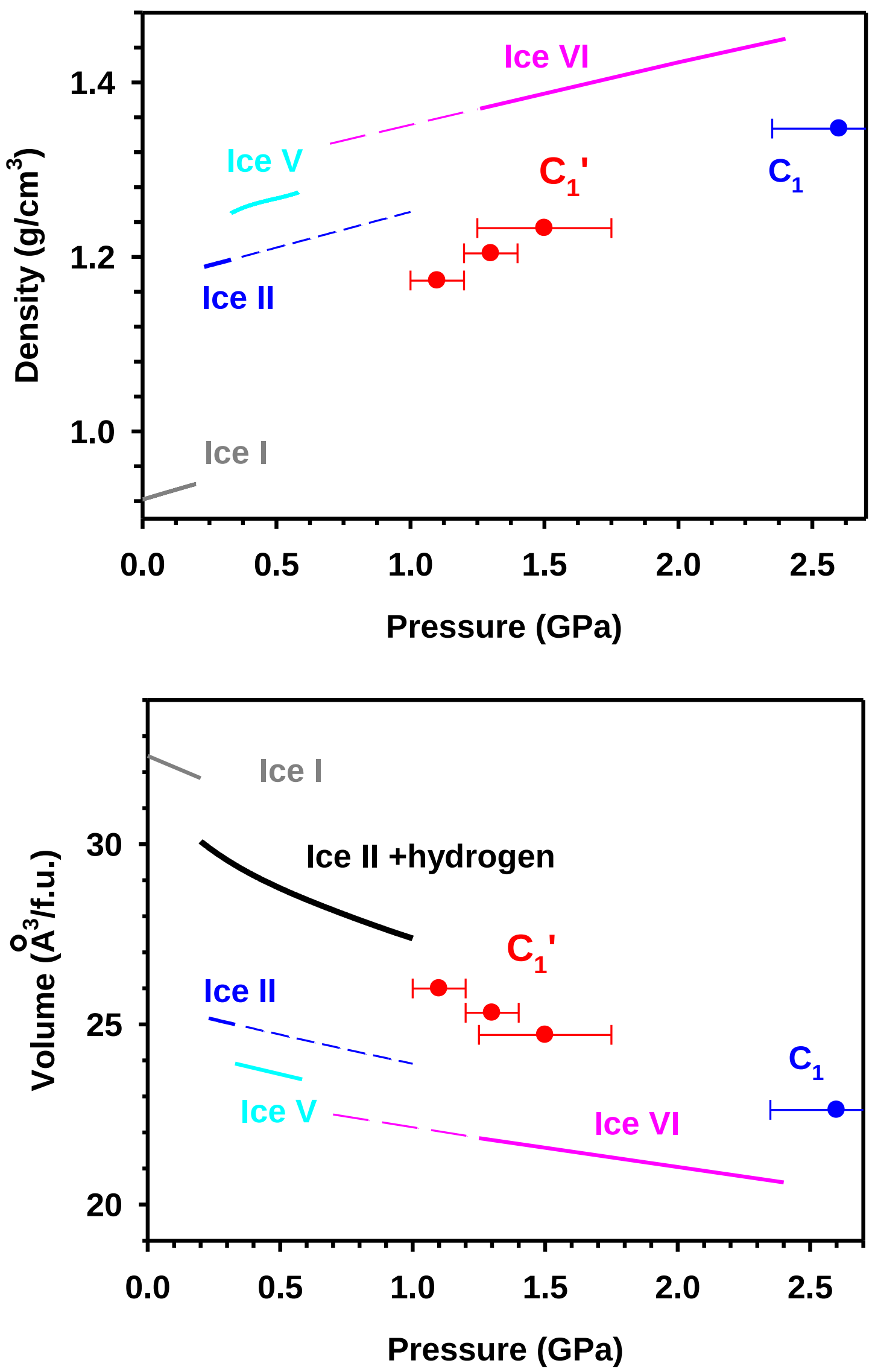

**Figure S5.** Density and specific volumes as a function of pressure of various ices and hydrogen clathrate hydrates studied here. The red filled circles and blue filled square are the data of this work for $C_1$' and $C_1$ phases, respectively. The error bar in the volume/pressure determination is smaller than the symbol size. The density (volume) data are from Refs. [16, 17] for ices I, II, and V, and ice VI, respectively. The solid lines correspond to the pressure range, where the data are collected, the dashed thin lines are linear extrapolation to the pressure range of interest. The black line represents a sum of the specific volumes of ice II and hydrogen, the data for the equation of state of solid hydrogen (phase I) are from Ref. [18].

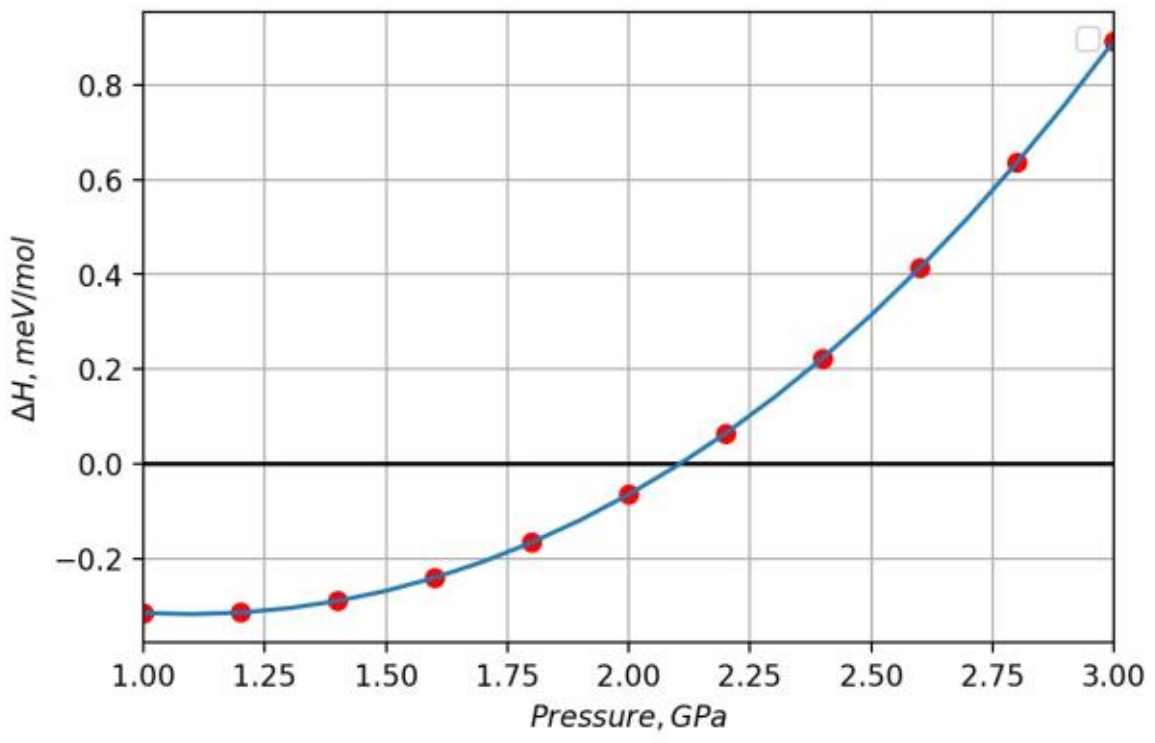

**Figure S6.** The enthalpy difference between $C_1$' and $C_1$ clathrate structures in the pressure range from 1 to 3 GPa. The ordered ***R3c*** structure was used to approximate $C_1$' phase (Table S2).

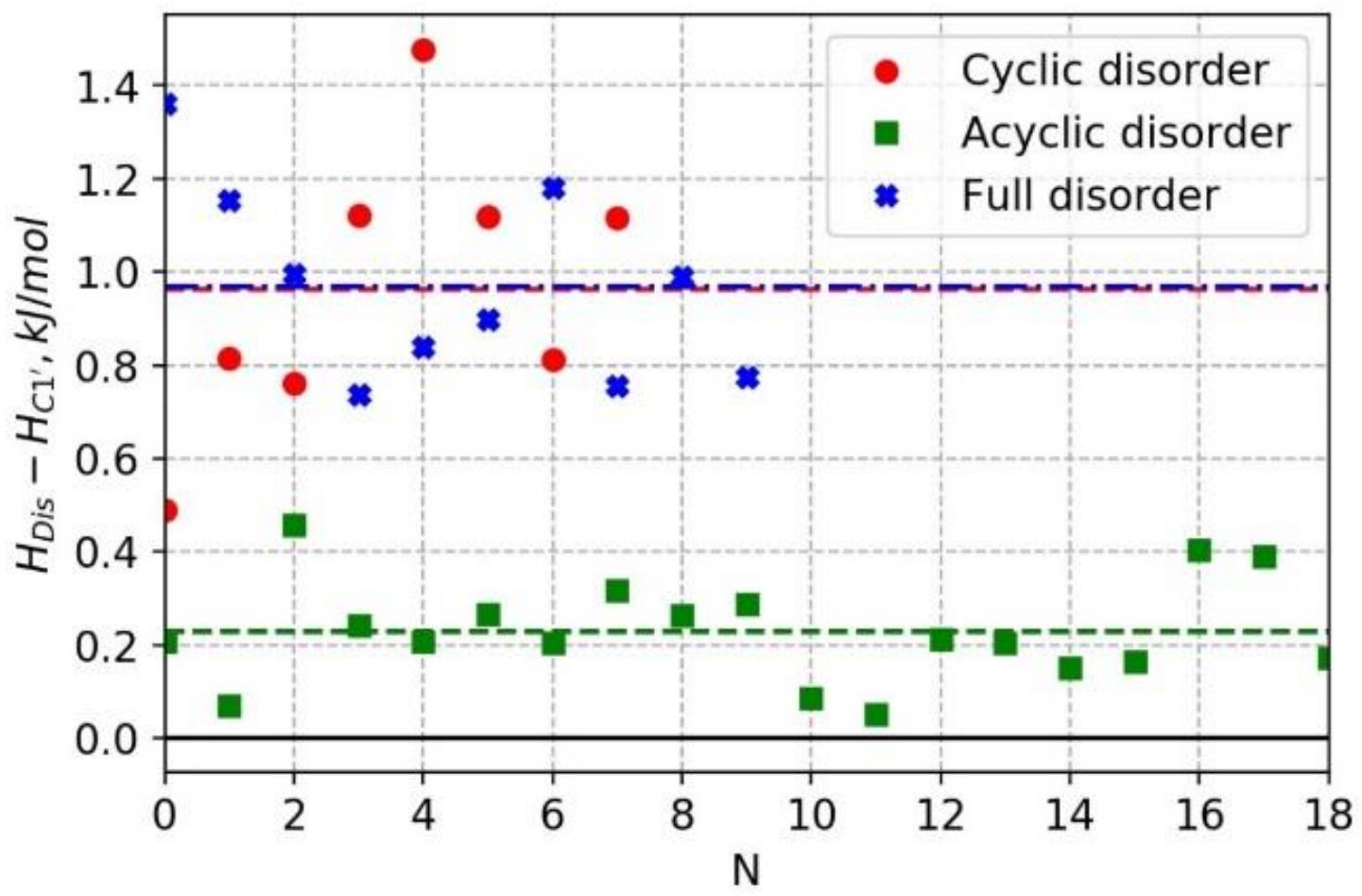

**Figure S7.** DFT computations of the difference between the average enthalpy of the series of disordered structures and the enthalpy of the USPEX-predicted structure for three hydrogen disorder types in the $(H_2O)_6 \bullet H_2$ crystal. N is a number of the particular ordered approximated structure.

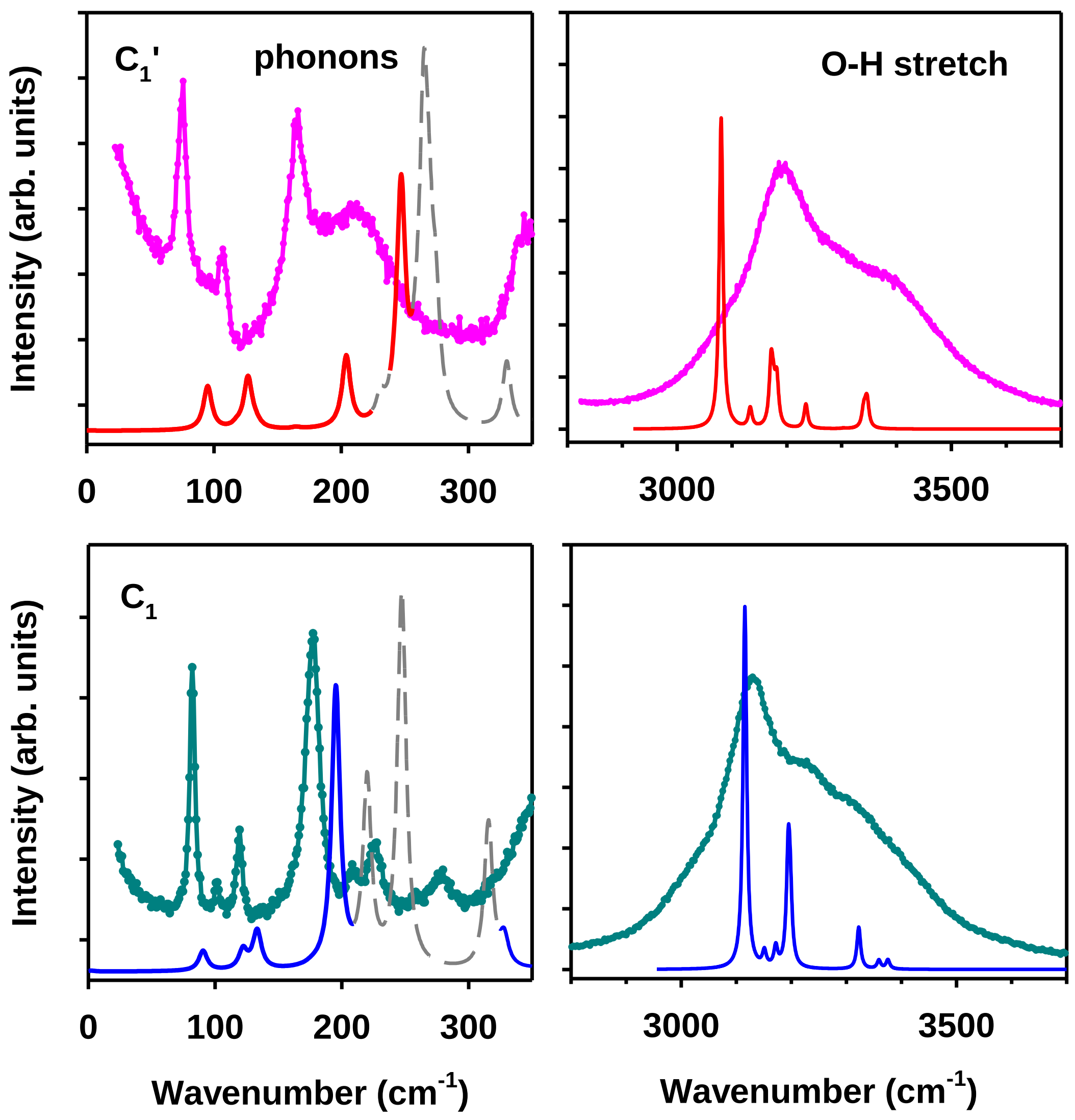

Figure S8. Raman spectra of $C_1'$ (at 1.2 GPa) and $C_1$ (at 2.3 GPa) clathrates in comparison to theoretical calculations (ordered approximation ***R3c*** for $C_1'$, Table S2) at 1.4 GPa. The rotational modes of $H_2$ molecules (libron) peaks calculated classically are shown by dashed gray lines. The calculated lines are broadened with a 4 $cm^{-1}$ line width (FWHM).

**Table S1.** Raman active modes of the $H_2O$ rings for two candidate $C_1$' and $C_1$ structures.

| Compound | Space group | Crystallographic positions | Raman modes per position | Total number |
|---|---|---|---|---|
| **$C_1$' (disordered)** | $\boldsymbol{R\bar{3}c}$, #167 | O1: 36f<br>H2-H5: 36f (population: 0.5) | $3A_{1g}+6E_g$ | 27 |
| **$C_1$' (ordered)** | $\boldsymbol{R3c}$, #161 | O1: 18b; O2: 18b<br>H2-H5: 18b | 3A+6E | 54 |
| **$C_1$** | $\boldsymbol{R\bar{3}}$, #148 | O1: 18f; O2: 18f<br>H2-H5: 18f | $3A_{1g}+3E_g$ | 36 |

## Table S2. Theoretical structure parameters

Ordered approximation of the $C_1$' (***R3c*** space group)

| Lattice parameters | Wyckoff position | x | y | z |
|---|---|---|---|---|
| | O (18b) | -0.140 | 0.299 | 0.179 |
| | O (18b) | 0.300 | -0.140 | 0.080 |
| | H (18b) | -0.120 | 0.318 | 0.019 |
| a = 12.650 Å | H (18b) | -0.226 | 0.229 | 0.184 |
| c = 6.047 Å | H (18b) | -0.062 | 0.240 | 0.324 |
| | H (18b) | 0.232 | -0.226 | 0.073 |
| | H (6a) | 0.000 | 0.000 | -0.329 |
| | H (6a) | 0.000 | 0.000 | -0.451 |

$C_1$ (***R-3*** space group)

| Lattice parameters | Wyckoff position | x | y | z |
|---|---|---|---|---|
| | O (18f) | -0.307 | 0.136 | 0.288 |
| | O (18f) | 0.102 | 0.476 | 0.144 |
| | H (18f) | -0.050 | -0.202 | 0.013 |
| a = 12.659 Å | H (18f) | -0.328 | 0.115 | 0.130 |
| c = 6.056 Å | H (18f) | 0.438 | -0.444 | 0.148 |
| | H (18f) | -0.427 | -0.402 | 0.228 |
| | H (6c) | 0.000 | 0.000 | 0.197 |
| | H (6c) | 0.000 | 0.000 | 0.319 |

**Table S3. The simulated shift of frequencies of the Raman spectrum upon decompression for the ordered approximant of the $C_1$' structure.**

| Modes # | Frequency, $cm^{-1}$ | | $\Delta F$, $cm^{-1}$ |
|---|---|---|---|
| | 1.4 GPa | 1.0 GPa | |
| 1 | 4405.14034 | 4409.504 | 4.36 |
| 2 | **3346.50593** | **3365.136** | **18.63** |
| 3 | **3344.06943** | **3360.897** | **16.83** |
| 4 | 3340.0981 | 3355.892 | 15.79 |
| 5 | 3277.74425 | 3282.39 | 4.65 |
| 6 | **3234.76376** | **3255.827** | **21.06** |
| 7 | 3181.67137 | 3214.492 | 32.82 |
| 8 | **3175.85037** | **3207.049** | **31.20** |
| 9 | **3171.7559** | **3202.887** | **31.13** |
| 10 | 3170.61579 | 3201.497 | 30.88 |
| 11 | 3133.20228 | 3164.022 | 30.82 |
| 12 | 3104.60894 | 3135.861 | 31.25 |
| 13 | 3103.00403 | 3131.109 | 28.10 |
| 14 | **3080.17171** | **3109.161** | **28.99** |
| 15 | 1730.93169 | 1728.487 | -2.44 |
| 16 | 1706.6902 | 1703.294 | -3.40 |
| 17 | 1687.54877 | 1685.878 | -1.67 |
| 18 | 1668.39872 | 1667.101 | -1.30 |
| 19 | 1653.21506 | 1652.671 | -0.54 |
| 20 | 1625.32577 | 1626.323 | 1.00 |
| 21 | 1616.85712 | 1618.439 | 1.58 |
| 22 | 1616.60489 | 1617.415 | 0.81 |
| 23 | 1014.51127 | 993.351 | -21.16 |
| 24 | 1004.89621 | 989.0379 | -15.86 |
| 25 | 982.58684 | 967.1265 | -15.46 |
| 26 | 971.4325 | 953.2301 | -18.20 |
| 27 | 915.09396 | 898.1604 | -16.93 |
| 28 | 873.82238 | 861.1575 | -12.66 |
| 29 | 873.36106 | 859.744 | -13.62 |
| 30 | 847.70824 | 833.3612 | -14.35 |
| 31 | 803.06359 | 790.4728 | -12.59 |
| 32 | 787.189 | 774.0914 | -13.10 |
| 33 | 770.64456 | 757.5792 | -13.07 |
| 34 | 738.70015 | 726.4568 | -12.24 |
| 35 | 710.28572 | 698.8609 | -11.42 |
| 36 | 706.15836 | 695.0396 | -11.12 |
| 37 | 682.78233 | 669.8434 | -12.94 |
| 38 | 612.14781 | 600.7006 | -11.45 |
| 39 | 606.41727 | 598.5988 | -7.82 |
| 40 | 601.15365 | 591.4758 | -9.68 |
| 41 | 384.05745 | 381.7588 | -2.30 |
| 42 | 365.33301 | 352.4292 | -12.90 |
| 43 | 330.15459 | 326.706 | -3.45 |
| 44 | 274.27119 | 264.7213 | -9.55 |
| 45 | 268.85321 | 261.5567 | -7.30 |
| 46 | 264.89407 | 260.5817 | -4.31 |
| 47 | 258.70337 | 259.7654 | 1.06 |
| 48 | 247.12061 | 239.6629 | -7.46 |
| 49 | **246.5146** | **239.6629** | **-6.85** |
| 50 | 243.80923 | 239.5291 | -4.28 |
| 51 | 230.80106 | 224.6122 | -6.19 |
| 52 | **203.9628** | **198.2805** | **-5.68** |
| 53 | **126.73870** | **128.78697** | **2.05** |
| 54 | **95.05041** | **97.57058** | **2.52** |

The Raman active modes are marked by bold red.